\documentclass[conference]{IEEEtran}
\IEEEoverridecommandlockouts
\usepackage{amsmath,amssymb,amsfonts}
\usepackage{graphicx}
\def\BibTeX{{\rm B\kern-.05em{\sc i\kern-.025em b}\kern-.08em
    T\kern-.1667em\lower.7ex\hbox{E}\kern-.125emX}}
\begin{document}

\title{An Optimized Signal Processing Pipeline for Syllable Detection and Speech Rate Estimation\\
{\footnotesize \textsuperscript{}}
\thanks{Supported by Visvesvaraya PhD scheme by Ministry of Electronics and Information Technology of India}
}
\author{\IEEEauthorblockN{Kamini Sabu, Syomantak Chaudhuri, Preeti Rao, Mahesh Patil}
\IEEEauthorblockA{\textit{Department of Electrical Engineering} \\
\textit{Indian Institute of Technology Bombay}\\
Mumbai, India \\
\{kaminisabu,syomantak,prao,mbpatil\}@ee.iitb.ac.in}
}

\maketitle

\begin{abstract}
Syllable detection is an important speech analysis task with applications in speech rate estimation, word segmentation, and automatic prosody detection. Based on the well-understood acoustic correlates of speech articulation, it has been realized by local peak picking on a frequency-weighted energy contour that represents vowel sonority. %The syllable detection performance however is critically dependent upon the choice of the speech analysis parameters and, in particular, the peak picking threshold. 
While several of the analysis parameters are set based on known speech signal properties, the selection of the frequency-weighting coefficients and peak-picking threshold typically involves heuristics, raising the possibility of data-based optimisation. In this work, we consider the optimization of the parameters based on the direct minimization of naturally arising task-specific objective functions. The resulting non-convex cost function is minimized using a population-based search algorithm to achieve a performance that exceeds previously published performance results on the same corpus using a relatively low amount of labeled data. Further, the optimisation of system parameters on a different corpus is shown to result in an explainable change in the optimal values. %Performance dependence on corpus, amount of training data and optimization cost function are discussed.
\end{abstract}

\begin{IEEEkeywords}
syllable detection, speech rate estimation, signal processing optimization, particle swarm optimization
\end{IEEEkeywords}

\section{\label{sec:intro} Introduction}

The syllable-level segmentation of speech is a critical component of speech processing systems with several applications including spoken language assessment for attributes such as speech rate and prosodic fluency. It has been addressed in the research literature over many decades, right up to the current day~\cite{1998morgan_fosler, 2000faltlhauser, Berisha2015, Chiranjeevi2018,rasanen2019}.  While a trained automatic speech recognition system can provide syllable- and word-level segmentations, it has been more common to employ far simpler speech signal processing pipelines to the same end without explicit phonetic segmentation. Further, given that most languages are characterized by syllabic structures comprising vowel nuclei, a method that computes the time variation of a vowel property such as sonority easily generalizes across languages. A vowel is articulated with a relatively open vocal tract configuration making it more energetic than consonants, especially in the vowel formant regions. Therefore performing peak and/or valley detections on a suitably computed band-weighted energy feature versus time can yield the number and locations of the syllables in an utterance. This is indeed the basis of the popular Mermelstein algorithm for syllable detection, widely deployed to this day \cite{Mermelstein1975,obin2013}. Speech rate is estimated by the number of detected syllables per second averaged over the desired utterance duration.

The performance of the automatic systems is evaluated by comparison with human annotated speech corpora. An algorithmic detection is considered correct if it falls within a tolerance window (in time) of a manually marked syllable nucleus. The following measures then serve to capture the system performance on a test dataset: achieved recall and precision of detections (further combined into F-score)~\cite{Chiranjeevi2018, 2009glass}, and the accuracy in speech rate estimation via the measured error in the total number of syllables detected over the utterance duration or the correlation between predicted and ground-truth speech rates %mean and standard deviation of speech rate error with respect to the ground-truth rate, and the correlation coefficient between the predicted and ground-truth speech rate
~\cite{Chiranjeevi2018, Berisha2015,jong2009}. 

A number of parameters are involved in the signal processing pipeline from the acoustic waveform of the utterance to the detected syllabic nuclei. The sub-modules involved are the short-time feature computation, temporal smoothing and, finally, picking the prominent peaks in the smoothened time series. The pipeline involves a set of parameters for each sub-module. A common feature computation is the weighted combination of band-wise short-time energies~\cite{Berisha2015,2007wang}. Reasonable choices for the processing parameters can be made in the case of short-time spectrum computation (i.e. window and hop sizes are based on speech signal stationarity), as also in the cut-off frequency of the smoothing filter applied to the sonority feature time series (based on the maximum expected syllable rate or fastest speaking speed).  The sub-bands are logarithmically spaced according to the auditory  model while the number of bands determines the dimensionality of the weight vector. On the other hand, the band-weighting coefficients as well as peak-picking threshold are typically selected heuristically. Such a methodology results in a completely unsupervised approach to syllable detection which, interestingly, has been reported to be competitive with recent deep learning approaches such as a single network of stacked BLSTMs trained on large multilingual corpus~\cite{landsiedel2011,rasanen2018,rasanen2019}.

The speech signal processing pipeline mentioned earlier is essentially a semantic abstraction based modular system, unlike neural network based learning systems where the time domain waveform or short-term spectra are input and speech-rate is output. The latter are expected, with their many more degrees of freedom, to require exponentially more data to train and may not generalize well to unseen conditions~\cite{2016shalev_arxiv_end2end}. We therefore explore here the approach followed by the majority of researchers and practitioners, in the context of the speech syllable detection task, of utilizing prior knowledge to build a pipeline of semantically meaningful sub-modules comprising various signal processing operations while considering further the joint optimization of a small set of critical parameters for given data characteristics.

Given that the choice of sub-module parameters has a significant impact on performance, there have been a few previous attempts at optimizing the parameters over a labeled dataset. Wang and Narayanan~\cite{2007wang} exhaustively search the parameter space to scan all possible local maxima of a performance measure in the given range where range and step-size are initialized by Monte-Carlo simulation and a parameter sensitivity analysis. Some optimization approaches avoid the need to explicitly determine the peak-picking threshold by resorting to statistical learning methods that directly learn the speech rate using GMM or HMM modeling~\cite{2000faltlhauser}.  Jiao et al.~\cite{Berisha2015} avoid the peak-picking threshold to set up a convex cost function for speech rate estimation involving integrating a special temporal density function to optimize the sub-bands' weighting coefficients for different dimensionalities of the coefficient vector. Yarra et al.~\cite{Chiranjeevi2018} take the approach of modeling the peak shape with a number of parameters and using peak classification (valid versus invalid) based on a feature vector of the estimated peak model parameter values. In a similar manner, Shankar and Venkataraman \cite{shankar2019} apply a loose threshold for syllable peak detection but focus on discriminating vowel peaks from consonant peaks using a trained classifier with several features computed on the peak.

In this paper, our aim is to carry out the joint optimization of the system parameters based on a task-specific cost function. We use an optimization technique that is applicable to convex as well as non-convex optimization problems. While learning algorithms optimize the (possibly intractable) task-specific performance measure indirectly via a surrogate loss function, optimization algorithms do a direct optimization\cite{Goodfellow-et-al-2016}.  We choose particle swarm optimization (PSO), a non-gradient based algorithm, as it removes the constraint of convexity on the cost function. This is important since the natural objective of our learning task is a non-convex function (such as, for example, the F-score for syllable detection). %The goal is to optimize the non-convex function directly since this can be expected to give us the best performance on a measure actually relevant to the end application. 
Other attractive features of PSO are its efficient implementation and its ability to find global optima in multidimensional spaces\cite{sengupta2018}. %\cite{ye2017_psoSGD}.

In the next section, we describe the system implementation and discuss the optimization of parameters. We present experimental results of the optimized system performance on the TIMIT test corpus \cite{TIMIT} in order to compare it with available previously published results of the similar systems with different approaches to parameter optimization. We also present and discuss the accuracies obtained on a children's oral reading corpus of interest to us in the context of reading fluency assessment~\cite{2018sabu_is_demo}.

%We aim at measuring speech rate in terms of syllables per second. The problem can be defined in three different ways - \#syllables prediction, syllable position prediction and syllable boundary prediction, later being the complex one predicting all the required info.  In this study, we focus only on the earlier i.e. \#syllables prediction which can also be simplified as predicting number of vowels since a syllable is defined to have a single vowel.

%Vowels are known to have energy concentrated in lower frequency bands. We compute energy in 4 major bands and weigh the same to get a weighed energy contour. The maxima in this contour are expected to correspond to the syllable positions and hence the count will give the number of syllables. To figure out the weights for each band, we use a particle swarm optimization (PSO) algorithm. Further, the peak-picking decision also affects the final performance. Therefore, PSO is trained to get the optimal peak-picking parameters.

%\subsection{Literature Review}

\section{Signal Processing Pipeline}
\subsection{Sub-modules and fixed parameters}
\begin{figure}[h]
  \centering
  \includegraphics[width=\linewidth]{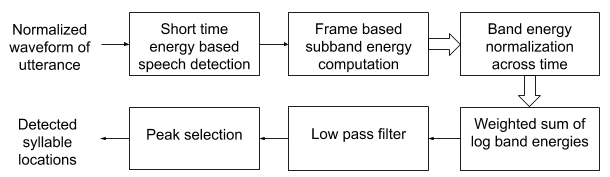}
  \caption{Signal processing pipeline for syllable detection}
  \label{fig:BD}
\end{figure}

As mentioned in Section~\ref{sec:intro}, the processing pipeline is essentially comprised of the computation of a temporal contour of weighted log sub-band energies followed by low-pass filter smoothing and subsequent peak picking. This is related to the band-wise correlation proposed in ~\cite{1998morgan_fosler} and adopted by several others including ~\cite{Berisha2015} and ~\cite{2007wang}. Figure~\ref{fig:BD} shows a block diagram of the processing chain. Based on the observation by Jiao et al.~\cite{Berisha2015} that 7 sub-bands, derived from the mel scale, provide performance similar to that obtained with 19 sub-bands assuming that the sub-band weighting coefficients are optimized in the same way, we use 7 logarithmically spaced sub-bands in our work. 

%to get good results provided their weighing factors are optimized using some optimization algorithms. They used convex optimization in~\cite{Berisha2015} and RNN in~\cite{Berisha2017}. Optimization can also be performed to select peaks with certain shape~\cite{Chiranjeevi2018}. In this work, we aim to jointly optimize the peak picking parameters along with the weighing factors. 

%To be moved... Very basic approach to speech rate estimation is counting number of syllables which appear as high energy peaks in short time energy contour. Instead of taking full band energy, one can just take the energy in certain bands and their correlation is observed to be highly following the vowel positions~\cite{1998morgan_fosler}. This concept has been used in~\cite{2007wang} where correlation is taken for energy contours of 19 subbands. They also added Gaussian filtering to remove spurious noisy peaks further followed by use of voicing property of vowels for peak validation. However, 19 subbands are too big a number and many of them may not be actually helping significantly. 

%Gaussian Mixture Models are used by~\cite{2000faltlhauser} to determine syllable positions using 12-dimensional MFCC features along with energy and zero crossing rate. 
%Rhythm based features can also be used for syllable detection where peaks in the envelope are counted~\cite{2009glass, Berisha2016}.

%The complete work flow for predicting number of syllables for given audio utterance involves feature computation followed by syllable prediction.

Figure~\ref{fig:BD} shows the different stages in the system to detect syllable nucleus locations from the given utterance waveform, normalized to [-1,1] range. All processing is carried out at 10 ms frame interval. A short-time energy based threshold is used to identify the speech frames. The short time window is 20 ms Hamming for the spectrum computation and subsequent sub-band energy computation.

In our work, we have considered the 7 frequency bands as provided in Table~\ref{tab:psoparam}. 
\begin{table}[ht]
  \caption{System parameters with search ranges for those that are jointly optimized by the PSO with the given settings}
  \label{tab:psoparam}
  \centering
  %\begin{ruledtabular}
  \begin{tabular}{ p{3cm} c }
    \textbf{Parameter} & \textbf{Value} \\
    \hline
    \hline
    Window size & 20 ms \\
    Frame size & 10 ms \\
    Speech energy threshold & -30 dB \\
    Smoothing filter cutoff & 7 Hz \\
    Sub-band edges (Hz) (7) & 60-370, 370-800,800-1400,\\
    &1400-2250,2250-3450, 3450-5130, 5130-7500 \\
    \hline
    Weights search range & -2.0 to 5.0 \\
    Peak prominence range & 0.01 to 10.0 \\
    \hline
    \hline
    \multicolumn{2}{c}{PSO settings} \\
    \hline
    No. of particles & 50    \\
    Maximum no. of iterations & 200    \\
    $\phi_p$, $\phi_g$, $\omega$ & 1.4962, 1.4962, 0.7298 \\
  \end{tabular}
  %\end{ruledtabular}
\end{table}
The subbands are defined by a trapezoidal frequency window with 50 Hz transition regions overlapping between bands. We thus obtain 7 band energy values corresponding to each 10 ms frame. The band energy contours so obtained are individually normalized across the utterance to a maximum value of 1.0 to account for loudness variability. Next, 7 band energy features are computed per frame by taking logarithm of band energies, after limiting very low values to a fixed small positive value.
% \begin{figure}[ht]
%   \centering
%   \includegraphics[width=\linewidth]{featcomp.png}
%   \caption{Feature computation.}
%   \label{fig:featcomp}
% \end{figure}
The log energy values are combined by a weighted sum in every frame to obtain a temporal envelope sampled at 10 ms as in equation~\ref{eq:weigh}.  %where the band-wise weighting coefficients are given by w1 through w7. 
\begin{equation}
E[n]=\sum_{i=1}^{7}{w_i e_i[n]}
\label{eq:weigh}
\end{equation}
where $e_i$[n] is the log energy feature in frame $n$ of band $i$ and $w_i$ is the corresponding weight.

 The temporal envelope so obtained is then smoothened with a low pass filter of cut-off frequency 7 Hz. The cut-off frequency is fixed from the knowledge of speech syllable rates as ranging up to around 7 syllables per sec~\cite{2017tivadar_JL_sylpersec}. The smoothened temporal envelope is searched for peaks (i.e. local maxima) that stand out in a local context. A widely accepted rule is to apply a threshold to each local maximum's ``prominence" value, defined as the distance from the nearest adjoining valley~\cite{2009glass, 2007wang}. The prominence threshold influences syllable detection performance directly. It is expected to depend on the data spectral characteristics, choice of sub-bands and the sub-band energy weighting coefficients. In this work, we investigate the joint optimization of the weighting coefficients, given a peak prominence threshold, based on a selected task-specific cost function as discussed next.

%Peak picking algorithm [give refs from related work e.g. Glass, not scipy] ~\cite{scipy_link} is applied on the smoothened contour to look for the vertical distance of a peak from its closes trough. In this case, peak prominence threshold also forms a feature to be optimized along with the weights.
 
% \begin{figure}[ht]
%   \centering
%   \includegraphics[width=\linewidth]{syllablePredict.png}
%   \caption{Predicting number of syllables.}
%   \label{fig:sylpredict}
% \end{figure}
\subsection{\label{sec:optimization} Cost functions for parameter optimization}

Performance measures in the context of syllable detection include comparisons between the number of human and automatically detected syllables at the utterance-level, and the precision and recall of automatic detection with reference to human detections. We note that a ``correct detection" is defined as that falling 
within an annotated vowel segment which itself is padded symmetrically to a minimum length of 50 ms if necessary~\cite{landsiedel2011}. An automatic detection that does not satisfy this criterion is considered a false alarm. If the end application requires speech rate estimation, on the other hand, the total number of detected syllables per utterance must be compared with the corresponding number of labeled syllables. The correlation coefficient between automatic and ground-truth speech rates (number of syllables/sec) is another useful performance measure provided the utterance duration is long enough for its reliable estimation~\cite{jong2009}.

Given the flexibility available with PSO in specifying the cost function computationally, we can exploit costs directly related to a performance measure that is eventually important to us in an application without worrying about differentiability. We therefore consider the two performance goals of syllable detection and speech rate estimation separately to obtain the two distinct cost functions below.

%The joint optimization of sub-module parameters can be achieved by using an optimization function corresponding to the performance measure that is eventually important to us in an application. Here we consider each of the syllable detection and speech rate estimation criteria separately.  

\begin{enumerate}

    \item F-score inverse (based on Precision and Recall in syllable detection)
    \begin{equation}
      cost = \frac{P + R}{2\ P\ R},
      \label{eq:cost2}
    \end{equation}
    where
    $$P = \frac{\text{\#correctly\ detected\ syl}}{\text{\#predicted\ syl}}$$
    $$R = \frac{\text{\#correctly\ detected\ syl}}{\text{\#actual\ syl}}$$
    
    \item Mean absolute error (MAE) for number of syllables per utterance
    \begin{equation}
      cost = \frac{\sum_{i=1}^{N}{| \#syl_{predicted} - \#syl_{actual} |}}{N},
      \label{eq:cost1}
    \end{equation}
    where $N$ = number of utterances.
    
\end{enumerate}
While the F-score is sensitive to the accuracy in syllable nucleus localization, the MAE relates to variation only in the total number of syllables detected over the utterance duration. %Given the presence of a peak thresholding step, we see that our optimization function is non-convex and does not provide the opportunity to use gradient-based search for optima. 

%Particle swarm optimization (PSO) gives us the flexibility to use a variety of optimization functions, to be minimized with respect to the 7 band-weighting coefficients and peak prominence threshold. The parameter settings used for feature computation and the PSO optimization are listed in Table~\ref{tab:psoparam}.

\section{Experimental Validation}
Comparing performances across different approaches in the literature is difficult in general due to differences in corpora and performance measures. We made our best attempt to identify comparable test scenarios and found published results on TIMIT test set with optimization, if any, carried out on the TIMIT train data set~\cite{Berisha2015, Chiranjeevi2018} .  We also provide results on a children's oral reading corpus created by us for the development of an automatic fluency assessment system. The corpus is named CS dataset (for `Campus School', where it was recorded).

\subsection{Datasets}
TIMIT dataset has 6300 sentences recorded in studio environment. Data is collected from 8 English dialect regions in U.S. and is phonetically transcribed~\cite{TIMIT}. The utterances have 4 to 20 syllables with duration ranging from 1 to 8 sec. The train and test sets are specified without speaker overlaps. The CS dataset comprises recordings from 10-14 year old Indian children reading aloud the text of short stories in English in quiet conditions. They have different levels of reading skills. The recordings are transcribed at word level. The phone level alignment is obtained next through forced alignment with a state of the art automatic speech recognition system~\cite{2017sabu_slate_lets}. The story recordings are chunked into sentence groups for the assessment of reading fluency including speech rate. The resulting utterance chunks range in duration between 0.5 s to 15 s. %have different number of syllables (1 to 35) with duration ranging from 0.5 to 15 sec. 
The train-test data distribution is given in Table~\ref{tab:data}. There is no speaker overlap between the train and test sets.

\begin{table}[th]
  \caption{Description of the data sets}
  \label{tab:data}
  \centering
  \begin{tabular}{ |c||c|c||c|c| }
    \hline
    \textbf{Dataset} & \multicolumn{2}{c||}{\textbf{CS data}} & \multicolumn{2}{c|}{\textbf{TIMIT data}} \\ \cline{2-5}
    & \textbf{Train} & \textbf{Test}  & \textbf{Train} & \textbf{Test} \\ \hline \hline
    No. of speakers & 16 & 7 & 462 & 168     \\ \hline
    No. of utterances & 2616 & 1202 & 4620 & 1680   \\ \hline
  \end{tabular}
  
\end{table}

\subsection{Parameter Optimization with PSO}
Particle Swarm Optimization (PSO) originally proposed by Kennedy and Eberhart in 1995~\cite{1995ebehart_icnn_pso} is widely considered a metaheuristic algorithm well suited to constrained non-convex optimization problems in high dimension space. It operates with a population of candidate solutions (``particles") moving in the specified parameter space, analogous to a swarm of birds. It has a low number of adjustable parameters. It does not require the cost function to be differentiable, or even to be specifiable by an equation. It is essentially applicable to \textit{any} cost function. We use the processing pipeline of Figure~\ref{fig:BD} with the desired performance measure to compute the cost or fitness function to be minimized in each iteration across the training data set in order to obtain the optimum 7-dimensional weight vector. We see, from equation~\ref{eq:weigh}, that the weights are continuous-valued scalars and that the optimal prominence threshold scales linearly with any constant scaling of the weights.  Table~\ref{tab:psoparam} lists the PSO parameter settings used in our work based on the common recommended values across diverse PSO applications~\cite{2000eberhart_ce_pso}. Constraints are placed on the parameters via the search ranges specified heuristically in Table~\ref{tab:psoparam}.

%proves to be better in terms of faster convergence and global minimum catching. This is due to many different randomly placed particles try to move towards minimum collectively.

\subsection{Evaluation of system performance}
A number of standard metrics are available to evaluate the performance of an automatic system on the tasks of syllable detection and speech-rate estimation on a given test dataset as listed below. %These are computed across all utterances in the test data set as follows~\cite{Berisha2015, Chiranjeevi2018}.
% values for theta were taken to be 0,1,1,0,0 with peak picking parameters 2,3,1. These values were chosen because log energies of the first two bands were found to be peaking at the desired peak positions. On the other hand, $3^{rd}$ and $4^{th}$ bands were observed to be peaking mostly at non-vowel positions. Further, peak prominence = 2 indicates the distance between a peak and its nearest trough. Here, it means that to be counted as valid peak, the peak should be at least 2 dB high from the nearest valley. Width of 3 points at relative height of 1 implies that at least 30 ms wider peak is expected at the position of computing peak prominence. Use of peak prominence in peak computations help overcome the issues due to reducing voicing energy across the utterance.

% As per \cite{Berisha2015}, number of features affect the system performance. They found that 7 mel frequency bands would give the best results. We compare the performance of our system when using 7 mel frequency sub-bands versus using 4 fixed bands corresponding to the expected 4 formant bands.

%Further, number of samples in the training set is known to act as trade off for performance and time complexity. This work studies the trade-off for the current system.

% Optimization can be joint or alternate.

%Performance evaluation is with following metrics -
\begin{itemize}
    \item Correlation coefficient between actual number of syllables and predicted number of syllables at utterance level~\cite{Berisha2015,rasanen2019}.
    %\item Mean and standard deviation of the error in predicted number of syllables in the utterance.
    \item SR error rate = Mean of the relative error in number of predicted syllables per utterance~\cite{Berisha2015}.
    %\item Mean and standard deviation of error in speech rate prediction.
    %\item Correlation coefficient between actual speech rate and predicted speech rate at utterance level.
    \item F-score for syllable detection~\cite{Chiranjeevi2018,obin2013,ludusan2016}.
\end{itemize}
We do not report speech rate correlations, expected to be unreliable due to the domination of short spurts  (i.e. utterances of duration $<$ 5 s) in the TIMIT dataset~\cite{jong2009}.

\section{Results and Discussion}
We present results based on the standard performance metrics provided in the previous section on each of the TIMIT and CS test data sets. The training data sets are used in the joint parameter optimization process. In particular, we study the variation in performance with different (i) training data size, (ii) cost functions.  Wherever possible, we provide a comparison of our system's performance with available published results on the TIMIT test set. We also investigate the possible corpus dependence of the optimal parameters by comparing the performance achieved on CS test data with parameters that are trained separately on each of the TIMIT and CS training data sets.%Table Y (3,4): We find that 7 bands are better than 4 bands.
%The proposed approach has been compared with other baseline methods and the performance measures can be seen from table~\ref{tab:results}. As expected, performance measures corresponding to the optimization function show great improvement, while others are comparable.

\begin{figure}[h]
  \centering
  \includegraphics[width=\linewidth]{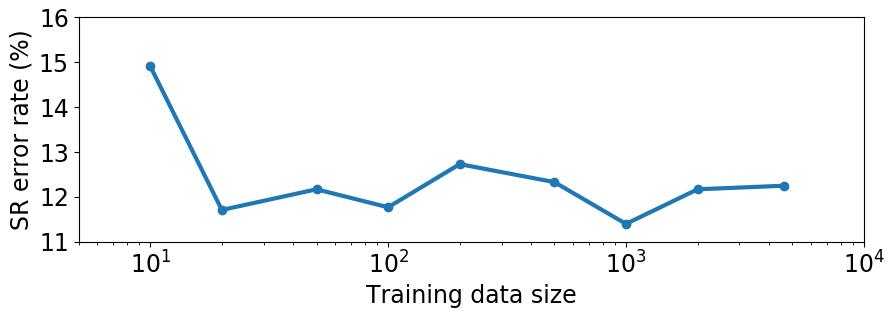}
  \caption{Performance on TIMIT test set using the 1/F-score cost function versus number of training sentences}
  \label{fig:trainsize}
\end{figure}

Figure~\ref{fig:trainsize} shows the variation of relative error in the number of syllables per utterance on the TIMIT test set with an increasing number of TIMIT train sentences used in the optimization, up to the maximum of 4620 sentences. We observe that the error decreases rapidly towards the 100 sentences mark and then fluctuates in a narrow band. This indicates that a relatively small data set is sufficient to achieve the joint parameter optimization by PSO. This is encouraging in view of the fact that phonetic (or syllable) labeled data sets are expensive to build, and therefore makes the optimization method more usable in practice than might be expected with a deep learning based network. The computation time for PSO optimization on the entire TIMIT training set (4620 sentences) with 100 iterations and 50 particles was limited to within 15 minutes on an i7 CPU with 16 GB RAM. The convergence at relative small training data set size was also observed by Jiao et al.~\cite{Berisha2015} in the context of a convex optimization of weight coefficients for the same task.   

% raining on different number of instances does not affect much. Even 100 sentences are found to be good enough to reach global minimum.

Table~\ref{tab:results} shows the results across performance metrics of the proposed system optimized with the two different cost functions as provided in Section~\ref{sec:optimization}. The F-score based cost function targets accurate detection (and localization) of individual syllables. The MAE, on the other hand, is the average error between the actual and predicted number of syllables per utterance, and is not so sensitive to the distinction between correct detections and false alarms. Since speech-rate depends only on the number of detected syllables, we expect MAE to be better suited to optimization for speech-rate estimation. Table~\ref{tab:results} also shows results from previous systems on the same task and same training-testing corpus. For a fair comparison with Yarra et al. \cite{Chiranjeevi2018}, we also provide results after a further post-processing step of rejecting peaks with low detected voicing within the parantheses in Table~\ref{tab:results}. 

% \begin{table*}[th]
%   \caption{Performance on TIMIT test data }
%   \label{tab:results}
%   \centering
%   \begin{tabular}{|c|c|c|c|c|c|c|c|}
%   \hline
%     \textbf{Method} & \textbf{Corr Coef} & \textbf{mean (S.D.)} & \textbf{relative error} & \textbf{mean (S.D.)} & \textbf{SR} & \textbf{F-score} \\
%     \textbf{} & \textbf{of \#syl} & \textbf{\#syl error} & \textbf{of \#syl (\%)}  & \textbf{SR error} & \textbf{corr} & \textbf{}\\ \hline \hline
%     Jiao et al. \cite{Berisha2015} & 0.842 & 1.57 (1.25) & 13.7 & 0.498 (0.39) & - & -  \\ \hline
%     Yarra et al. \cite{Chiranjeevi2018} & - & - & - & - & 0.693 & 84.16    \\ \hline
%     % Proposed \\ \hline
%     Proposed (1/F-score) & 0.906 & 1.47 (1.25) & 11.4 & 0.496 (0.424) & 0.678 & 87.00     \\ \hline
%     Proposed (MAE) & 0.890 & 1.34 (1.21) & 10.8 & 0.448 (0.392) & 0.639 & 85.40     \\ \hline
%   \end{tabular}
% \end{table*}

\begin{table}[th]
  \caption{Performance on TIMIT test data with model trained on 1000 training instances of TIMIT. Performance with voicing applied shown in parentheses.}
  \label{tab:results}
  \centering
  \begin{tabular}{|c|c|c|c|}
  \hline
    \textbf{Method} & \textbf{Corr Coef} & \textbf{SR error} & \textbf{F-score} \\
    \textbf{} & \textbf{of \#syl} & \textbf{rate (\%)}  & \textbf{}\\ \hline \hline
    Jiao et al. \cite{Berisha2015} & 0.842 & 13.7 & -  \\ \hline
    Yarra et al. \cite{Chiranjeevi2018} & - & - & 84.16    \\ \hline
    % Proposed \\ \hline
    Proposed (1/F-score) & 0.906 (0.917) & 11.4 (11.89) & 87.73 (88.29)   \\ \hline
    Proposed (MAE) & 0.890 (0.913) & 10.8 (9.94) & 86.15 (87.88)     \\ \hline
  \end{tabular}
\end{table}
We observe from Table~\ref{tab:results} that the F-score achieved by the proposed system using either cost function is superior to the previously published result based on peak classification~\cite{Chiranjeevi2018}.  We further note that the proposed joint optimization exceeds the performance of Jiao et al.~\cite{Berisha2015} (errors going down, correlations going up) on all the performance metrics with both cost functions. As expected, the F-score based cost function also provides better F-score performance while the MAE cost function is better for speech rate estimation.
This validates our hypothesis about the benefit of employing a task-relevant cost function for the optimization of system parameters. Further, the analysis of observed syllable detection errors shows that these typically come from short duration vowels (as syllabic nuclei) and syllables devoid of vowels (so that there is no clear valley in the temporal envelope between syllables).

\begin{table}[ht]
  \caption{Performance on CS test data with parameters optimized on 1000 TIMIT sentences, or on 1000 CS train utterances, both with (1/F-score) as cost function.}
  \label{tab:CSresults}
  \centering
  \begin{tabular}{|c|c|c|c|}
  \hline
    \textbf{Training} & \textbf{Corr Coef} & \textbf{SR error} & \textbf{F-score} \\
    \textbf{corpus} & \textbf{of \#syl} & \textbf{rate (\%)}  & \textbf{} \\ \hline \hline
    TIMIT & 0.959 & 9.9 & 87.56 \\ \hline 
    CS & 0.980 & 7.3 & 93.74 \\ \hline 
  \end{tabular}
\end{table}

\begin{figure}[ht]
  \centering
  \includegraphics[width=\linewidth]{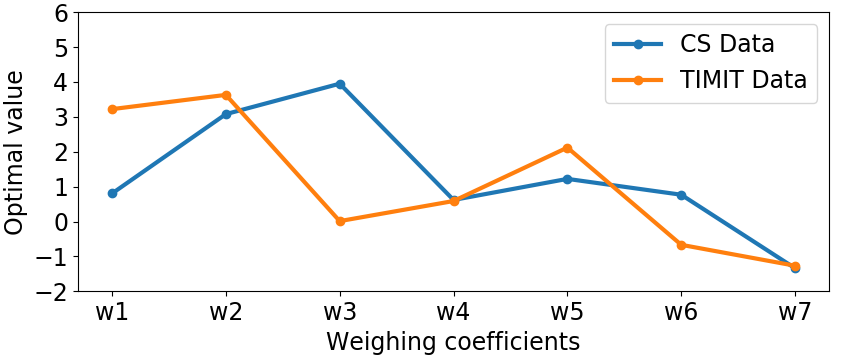}
  \caption{Optimal weight values obtained by training with 1/F-score cost function on 1000 training utterances of TIMIT data and on CS data. Peak prominence threshold is 2.416 for TIMIT data and 7.341 for CS data.}
  \label{fig:featimp}
\end{figure}

Table~\ref{tab:CSresults} presents results on the CS test data set. With the TIMIT train set used for optimization, we observe better performance on all metrics relative to those obtained on the TIMIT test set as reported in Table~\ref{tab:results}. This may be explained by the generally slower speech rate and clearer articulation by children reading aloud. But, more importantly, we observe the superiority of parameters optimized on training data drawn from the same corpus. This demonstrates the corpus dependence of the optimal parameters. This aspect has been noted by Jiao et al.~\cite{Berisha2015} as well. While the lack of generalization may appear to be a drawback, we keep in mind our previous observation that the amount of data required for optimization to converge is relatively low for any given corpus, making data-based optimization practical in many scenarios. A better understanding of parameter dependence on corpus comes from comparing the optimal sub-band weights
determined across the two corpora in Figure~\ref{fig:featimp} . We see that the shapes of the contours are similar with the CS corpus weights being more skewed to the right. 
This frequency scaling may be explained by the fact that the CS data is characterized by higher formant frequencies due to the shorter vocal tract lengths of children.  We also note that the weights are smaller in both corpora for the higher frequency bands with negative values in the highest band. This suggests that the features which make syllable nuclei prominent are, both, the strong presence of vowel formant band energies and the simultaneous absence of high-frequency energy concentration that characterizes consonants such as fricatives and stops. 

\section{Conclusion}
We have considered the task of syllable detection in speech signals using different measures of performance. We have demonstrated the data-based optimization of a popular signal processing pipeline using an efficient method, in terms of both computation time and labeled training data size, that places no constraints on the cost function form. Consistent improvements are observed on performance measures of interest based on the appropriate choice of cost function. Future work targets extensions to different language corpora as also to further improvements in performance by incorporating new post-processing methods on the detected peaks.

\bibliographystyle{IEEEtran}
\bibliography{mybib}

\end{document}